\documentstyle[sprocl,epsfig]{article} 

\renewcommand{\mathrm}[1]{{\rm #1}} 
\newcommand{\keV}{\,\mathrm{ke\kern-1pt V}}
\newcommand{\MeV}{\,\mathrm{Me\kern-1pt V}}
\newcommand{\GeV}{\,\mathrm{Ge\kern-1pt V}}
\newcommand{\EGeV}{E\,\mathrm{[Ge\kern-1pt V]}}
\newcommand{\TeV}{\,\mathrm{Te\kern-1pt V}}

\newcommand{\pom}{{\rm I\! P}} \newcommand{\reg}{{\rm I\! R}}
\newcommand{\sgsp}{\sigma_\mathrm{tot}^{\gamma^*p}}

\newcommand{\sgp}{\sigma_\mathrm{tot}^{\gamma p}}

\newcommand{\ft}{F_2} \newcommand{\ftxq}{F_2(x,Q^2)}
\newcommand{\fl}{F_L}

 \newcommand{\zpc}[1]{Z.\
Phys.\ {\bf C #1}}

\newcommand{\plb}[1]{Phys.\ Lett.\ {\bf B #1}}
\newcommand{\npb}[1]{Nucl.\ Phys.\ {\bf B #1}}
\newcommand{\epjc}[1]{Eur.\ Phys.\ J.\ {\bf C #1}}

\newcommand{\sjnp}[1]{Sov.\ J.\ Nucl.\ Phys.\ {\bf #1}}
\newcommand{\spj}[1]{Sov.\ Phys.\ JETP {\bf #1}}

  \newcommand{\dy}{DESY Report }

\begin{document}

\title{STATUS OF\\ MEASUREMENTS AND INTERPRETATION OF\\ TOTAL REAL AND
VIRTUAL\\ PHOTON-PROTON CROSS SECTIONS}

\author{CH. AMELUNG\\ (for the ZEUS and H1 Collaborations)\\[1ex]}

\address{Physikalisches Institut der Universit\"at Bonn, Nu{\ss}allee
12,\\53115 Bonn, Germany\\E-mail: amelung@physik.uni-bonn.de\\} 

\maketitle\abstracts{The status of measurements and interpretation of
the proton structure function $\ft$ is summarized. The measurements
are subjected to DGLAP fits, from which the gluon density is extracted
and a comparison to measurements of $\ft^c$ is performed. The
longitudinal structure function, $\fl$, is extracted via an
extrapolation of $\ft$. At low $Q^2$, the transition from virtual to
real photon exchange is studied.}

\section{Introduction}

The proton structure functions carry information about the dynamics of
quarks and gluons inside the proton, and they can be used to test the
theory of the strong interaction. The measurement of the structure
functions is an important part of the physics program at the $ep$
collider HERA, and has been performed by the ZEUS and H1
collaborations.

\subsection{Kinematics of Deep Inelastic Scattering}

Inelastic positron-proton scattering, $e^+p\to e^+X$, can be described
in terms of two kinematic variables, $x$ and $Q^2$, where $x$ is the
Bjorken scaling variable and $Q^2$ the negative squared four-momentum
transfer. In the absence of initial- and final-state radiation,
$Q^2=-q^2=-(k-k')^2$ and $x=Q^2/(2P\cdot q)$, where $k$ and $P$ are
the four-momenta of the incoming positron and proton, respectively,
and $k'$ is the four-momentum of the scattered positron.  The
fractional energy transfer to the proton in its rest frame, $y$, can
be related to $x$ and $Q^2$ by $Q^2=sxy$, where $s=4E_eE_p$ is the
square of the positron-proton center-of-mass energy. Here, $E_e$
($27.5\GeV$) and $E_p$ ($820\GeV$) are the positron and proton beam
energies, respectively.  For $Q^2 \ll M_Z^2$, positron-proton
scattering is characterized by the exchange of a photon. The
photon-proton center-of-mass energy squared is defined as
$W^2=Q^2(1-x)/x+m_p^2$.

The kinematic variables are closely related to the energy, $E_e'$, and
angle, $\theta$, of the scattered positron, where $\theta$ is measured
with respect to the positron beam direction: $y=1-{E_e'}/{2E_e}(1+\cos
\theta)$ and $Q^2=2E_eE_e'(1-\cos\theta)$.

\begin{figure}[tb]

\begin{minipage}{6cm}

\epsfig{figure=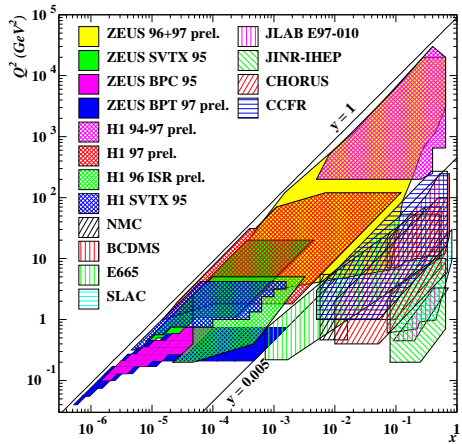,width=6cm}


\end{minipage}\hfill\begin{minipage}{5.5cm}\caption{\label{ARNIE} The
range in the kinematic plane $Q^2$ vs.\ $x$ covered by the HERA
experiments and by various fixed-target experiments. In most parts of
the $Q^2$ range, ZEUS and H1 data overlap with the fixed-target data.
At low $Q^2$, the transition from $\gamma^*p$ to $\gamma p$ scattering
can be studied. Measurements at medium $Q^2$ are used for studies and
tests of pQCD. At high $Q^2$, electroweak effects become important and
searches for new physics are performed.}\end{minipage}

\end{figure}

HERA has extended the previously accessible range in the kinematic
variables $x$ and $Q^2$ by two orders of magnitude, as shown in Fig.\
\ref{ARNIE}.

\subsection{Cross Sections and Structure Functions}

The double-differential cross section for inelastic $e^\pm p\to e^\pm
X$ scattering can be expressed in terms of structure functions as
\begin{equation}\label{DSIGMA}\frac{d^2\sigma(e^\pm p\to e^\pm
X)}{dy\,dQ^2} =\frac{2\pi\alpha^2}{yQ^4} \left(Y_+\ft-y^2\fl\mp
Y_-xF_3\right)(1+\delta_r),\end{equation} where $Y_\pm=1\pm (1-y)^2$,
and the radiative correction to the Born cross section, $\delta_r$, is
a function of $y$ and $Q^2$, but to a good approximation independent
of $F_2$ and $F_L$. From the measured quantity, the
double-differential $ep$ cross section, $\ft$ is extracted by making
suitable assumptions about $\fl$ and $xF_3$.

At $Q^2\ll M_Z^2$, where $xF_3$ can be neglected, the $ep$ cross
section can also be expressed in terms of total cross sections for the
absorption of transversely and longitudinally polarized virtual
photons, \begin{equation}\frac{d^2\sigma(e^\pm p\to e^\pm X)}{dy\,dQ^2}
= \Gamma \left( \sigma_T^{\gamma^* p} + \epsilon\, \sigma_L^{\gamma^*
p}\right) (1+\delta_r),\end{equation} where $\Gamma = \alpha Y_+/(2\pi
y Q^2)$, and $\epsilon= 2(1-y)/Y_+$. At low $x$, the photon-proton
cross sections are related to the structure functions by
\begin{equation}\label{FTWOSIGTOT}\sgsp=\sigma_T^{\gamma^* p}
+\sigma_L^{\gamma^* p}\approx
{4\pi^2\alpha}/{Q^2}\cdot\ft\quad\mathrm{and}\quad \sigma_L^{\gamma^*
p}\approx {4\pi^2\alpha}/{Q^2}\cdot\fl.\end{equation}

The so-called reduced cross section \begin{equation}\sigma_r=
\ft-\frac{y^2}{Y_+}\fl\mp \frac{Y_-}{Y_+}xF_3
\end{equation} can also be defined, which differs from the
double-differential cross section only by a kinematic factor, and can
thus be extracted without assumptions about $\fl$ and $xF_3$. In most
of the kinematic region covered by the present analyses, $y^2\fl$ and
$Y_-xF_3$ are negligible, and thus $\sigma_r\approx\ft$.

In lowest-order QCD, $\ft$ can be expressed as the charge-weighted sum
of quark and antiquark densities in the proton. While fixed-target
experiments probe predominantly the valence-quark content of the
proton, the HERA experiments are mainly sensitive to the sea
contribution arising from gluon radiation off quarks and subsequent
gluon splitting into quark-antiquark pairs. Only at very high $Q^2$ do
the HERA experiments also probe the valence quarks.

\section{\boldmath Structure Functions at High and Medium $Q^2$}

The measurement of $\ft$ is an inclusive cross-section measurement
using the angle and energy of the detected scattered positron. Good
detection efficiency and purity require both tracking and calorimetric
energy measurements. The ZEUS $\ft$ measurement at medium and high
$Q^2$ used the uranium-scintillator calorimeter and the central
tracking detector. The acceptance of the latter, however, is limited
to $16^\mathrm{o}<\theta<165^\mathrm{o}$. The H1 collaboration
upgraded their detector in 1997 to improve the positron detection at
small scattering angles by adding four rear discs to their microvertex
detector (Backward Silicon Tracker, BST), which extend the tracking
acceptance down to $3^\mathrm{o}$.

\subsection{Measurement of $\ft$}

\begin{figure}[tb] 

\begin{minipage}{6cm}

\epsfig{figure=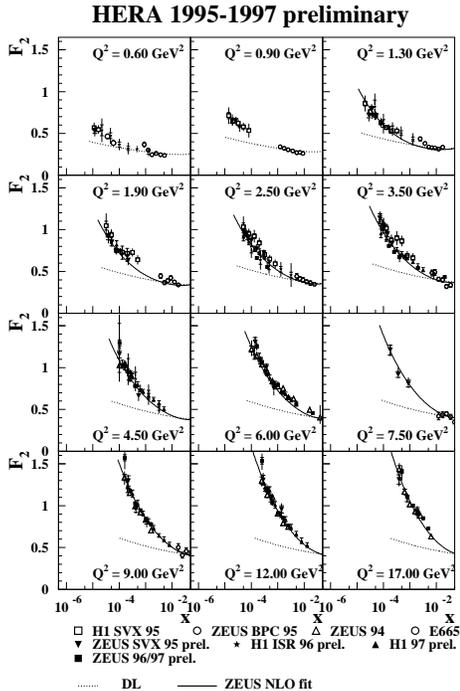,width=6cm}


\end{minipage}\hfill\begin{minipage}{5.5cm}\caption{\label{HERAF2}
Overview of $\ft$ measurements at low and medium $Q^2$. The symbols
denote measurements from ZEUS, H1 and E665 in different parts of the
kinematic range and by special analysis techniques (SVX: shifted
vertex, ISR: events with initial state radiation, BPC: using a special
calorimeter close to the beam line). The solid line is the result of a
DGLAP fit, the dotted line a Regge-type parameterization by Donnachie
and Landshoff \protect\cite{DL94}.}\end{minipage}

\end{figure}

Figure \ref{HERAF2} shows an overview of $\ft$ measurements at low and
medium $Q^2$. In selected kinematic regions, ZEUS and H1 have achieved
$\ft$ results with a precision of 1\% (stat.) and 3\% (sys.),
comparable to the typical precision of fixed-target experiments.

The measured $\ft$ can be well described by a form $\ft\propto
x^{-\lambda}$. The quantity $\lambda$ decreases as $Q^2$ decreases,
corresponding to the rise of $\ft$ becoming less steep. As can be seen
from Eq.\ (\ref{FTWOSIGTOT}), the rise of $\ft$ with $x$ corresponds
to an energy dependence of $\sgsp$ of $\sgsp\propto W^{2\lambda}$.

Two alternative and complementary theoretical frameworks are used to
interpret these data. At medium to high $Q^2$, i.e.\ in hard
interactions, pQCD is applicable, because $\alpha_s$ is small. At low
$Q^2$, i.e.\ in soft interactions, $\alpha_s$ is large, and
perturbative calculation techniques break down. This is the domain of
Regge theory.

\subsection{DGLAP Fits}

\begin{figure}[tb] 

\begin{minipage}{5.9cm}

\epsfig{figure=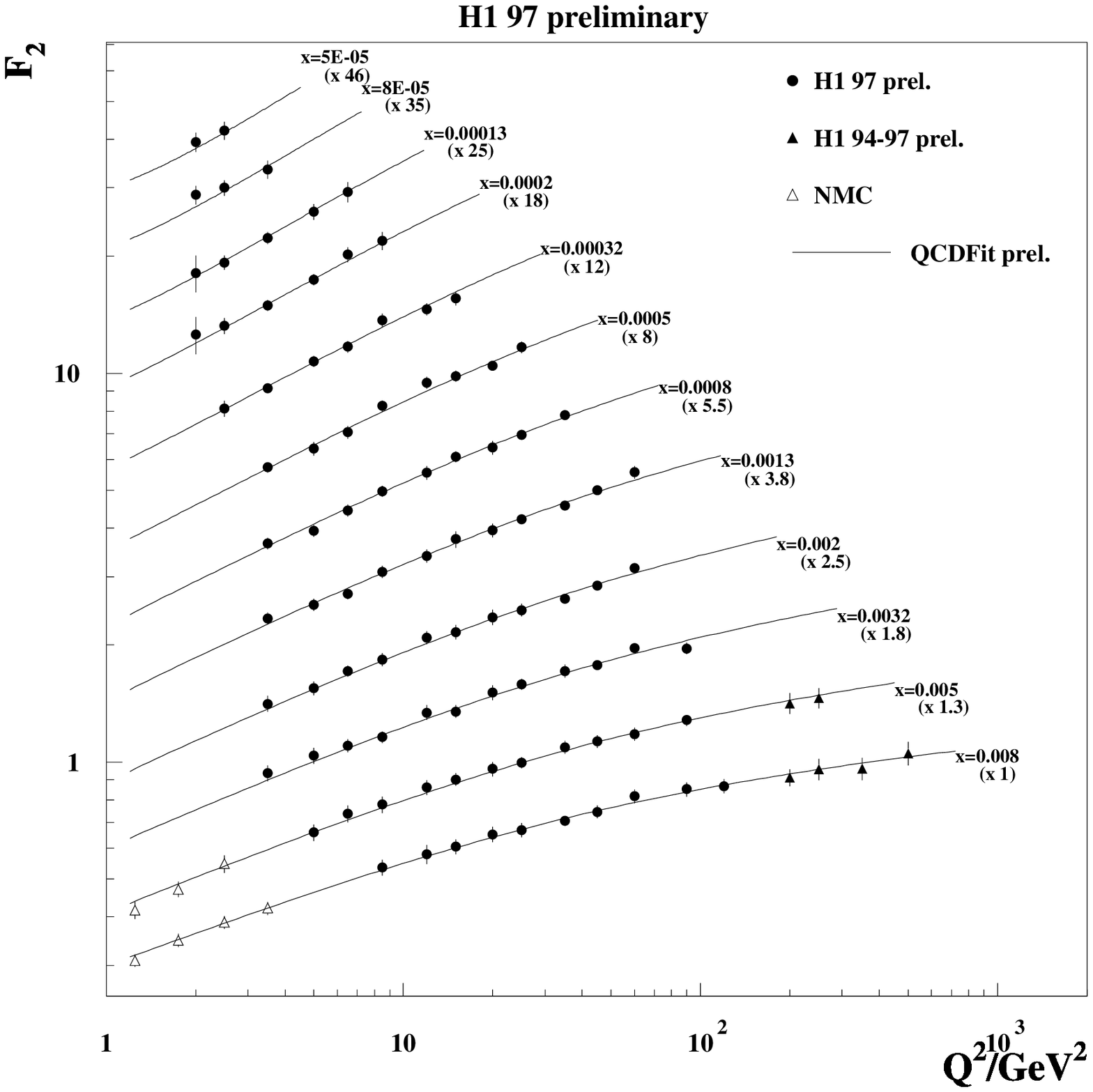,width=5.9cm} 


\end{minipage}\hfill\begin{minipage}{5.9cm}

\epsfig{figure=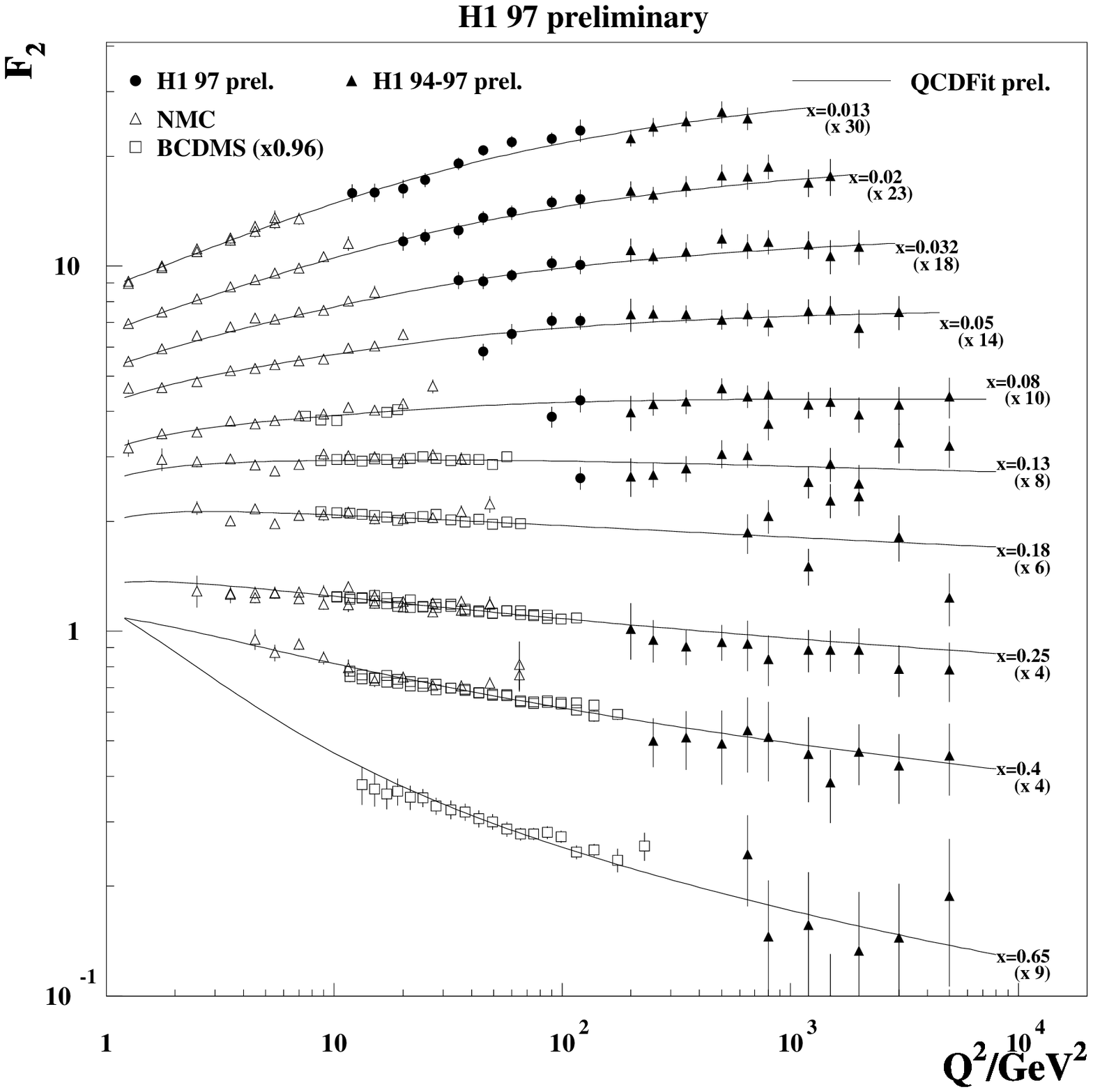,width=5.9cm}

\end{minipage}

\caption{\label{H1SCALING}The scaling violations of $\ft$ as measured
by H1 and the fixed-target experiments NMC and BCDMS, displayed as
$\ft$ vs.\ $Q^2$. The solid line is the result of a DGLAP fit, which
is able to describe $\ft$ over four orders of magnitude in $Q^2$.}

\end{figure}

The DGLAP equations \cite{DGLAP} predict the evolution of parton
densities with $Q^2$ (but make no statement about the $x$ dependence).
They relate the derivative of gluon and quark densities with $Q^2$ to
the convolution of gluon and quark densities with the splitting
functions $P_{ij}$. The latter are calculable in pQCD and are a
measure of the probability to find a parton $i$ radiated by a parton
$j$ as a function of $Q^2$.

In order to compare the $Q^2$ evolution of parton densities as
predicted by the DGLAP equations to that of the measured $\ft$ data,
the gluon, sea quark and valence quark densities are parameterized by
a suitable form at a starting scale $Q^2_0$. The parton densities are
evolved to other values of $Q^2$, and $\ft$ is calculated from the
quark densities. The free parameters of the parton density
parameterizations are adjusted such that the resulting $\ft$
parameterization fits the data above a cut-off value of
$Q^2_\mathrm{min}$. 


Figure \ref{H1SCALING} shows the result of such a fit to $\ft$ data
from H1. The value of $\ft$ increases at small $x$ when increasing
$Q^2$, while it decreases simultaneously at large $x$. The latter is a
consequence of valence quarks radiating gluons and thus losing
momentum, while the former results from the radiated gluons splitting
into low-momentum quark-antiquark pairs. Scaling, i.e.\
$Q^2$-independence of $\ft$, can be seen only at $x\approx 0.1$. The
fit describes the measured $\ft$ values well over four orders of
magnitude in $Q^2$, which is an impressive confirmation of QCD. The
fit does not deviate from the data down to the (surprisingly low)
value of $Q^2=1\GeV^2$, where the application of pQCD is somewhat
questionable.

\subsection{The Gluon Density}

\begin{figure}[tb] 

\begin{minipage}{6cm}

\epsfig{figure=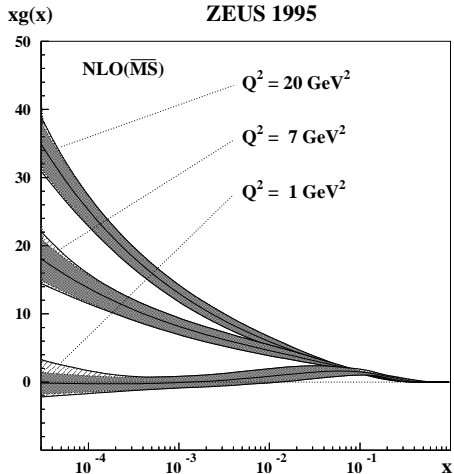,width=6cm} 


\end{minipage}\hfill\begin{minipage}{5.5cm}\caption{\label{GLUON}The
gluon density resulting from a DGLAP fit to $\ft$ data from ZEUS, for
three different values of $Q^2$. The shaded bands show the uncertainty
of $xg(x,Q^2)$. At $Q^2=1\GeV^2$, the gluon density as extracted from
the fit becomes compatible with zero at $x<10^{-2}$.}\end{minipage}

\end{figure}

The gluon density can be extracted from a DGLAP fit to measured $\ft$
data. The result from ZEUS is shown in Fig.\ \ref{GLUON}. Much effort
has been put into a reliable determination of the precision of this
extraction, which is 10\%--15\%. A comparison of the value of
$xg(x,Q^2)$ to the value of $\ftxq$ at low $x$ shows that the
overwhelming fraction of partons in the proton at low $x$ are gluons.

Like the quark densities, the gluon density increases with $Q^2$.
However, when going down to $Q^2$ values as low as $1\GeV^2$, the
gluon density from the fit becomes compatible with zero at
$x<10^{-2}$, while the sea quark density remains positive and rises up
to the smallest accessible values of $x$. This seems to be in
contradiction to the concept that quarks at low $x$ are produced by
gluon splitting, and may indicate a problem in applying pQCD at such
low values of $Q^2$.

\subsection{Measurement of $\ft^c$}

\begin{figure}[tb] 

\begin{minipage}{6cm}

\epsfig{figure=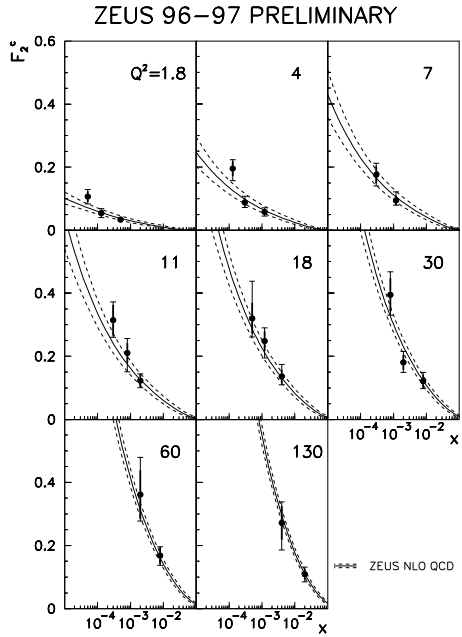,width=6cm}


\end{minipage}\hfill\begin{minipage}{5.5cm} \caption{\label{ZEUSFC}
The charm structure function $\ft^c$ as a function of $x$ in bins of
$Q^2$, measured by ZEUS. The solid line shows the central value of
$\ft^c$ as calculated from a DGLAP fit to $\ft$ data, which did not
include the $\ft^c$ measurements shown here. The dashed lines indicate
the uncertainty of the fit result. In principle, the observed
agreement of the measured $\ft^c$ with that extracted from a DGLAP fit
to the inclusive $\ft$ is an important consistency check of the DGLAP
formalism; however, the significance of this observation is reduced
here due to the extrapolation technique making use of the measured
inclusive $\ft$.}\end{minipage}

\end{figure}

The charm structure functions $F_i^c$ are defined in analogy to
(\ref{DSIGMA}) with the requirement that the final state contains a
$c\bar{c}$ pair: \begin{equation}\label{DSIGMAC}\frac{d^2\sigma(e^\pm
p\to e^\pm c\bar{c}X)}{dy\,dQ^2} =\frac{2\pi\alpha^2}{yQ^4}
\left(Y_+\ft^c-y^2\fl^c\mp Y_-xF_3^c\right)(1+\delta_r).\end{equation}
At HERA, charm is generated predominantly in the photon-gluon fusion
process $\gamma g\to c\bar{c}$, and is detected by reconstructing the
fragmentation products of the $c$ quark, usually in the channel
$D^*\to D^0\pi_\mathrm{slow}$, where the $D^0$ decays further to
$K\pi$, $K\pi\pi\pi$, or semi-leptonically. At the current level of
precision, $\fl^c$ and $F_3^c$ can safely be neglected. In contrast to
the fully inclusive $\ft$ measurements, further input is needed, e.g.\
the fragmentation function for $c\to D^*$ and the $D^*\to D^0\pi$
branching ratio. In addition, the event selection efficiency is quite
low, because experimentally only a limited kinematic region of the
$D^*$ is accessible. Therefore, a Monte Carlo simulation (based on the
measured inclusive $\ft$) has to be used to extrapolate to the full
kinematic region.

The result of the extraction of $\ft^c$ by ZEUS is shown in Fig.\
\ref{ZEUSFC}. The charm structure function $\ft^c$ exhibits a steeper
rise toward low $x$ than the fully inclusive $\ft$, and stronger
scaling violations. The charm content of the proton, $\ft^c/\ft$,
rises from about 10\% at $Q^2=1.8\GeV^2$ to 25\% at $130\GeV^2$.

\subsection{Extraction of $\fl$}

\begin{figure}[tb] 

\begin{minipage}{6cm}

\epsfig{figure=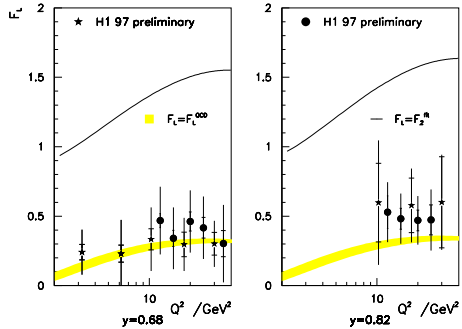,width=6cm}


\end{minipage}\hfill\begin{minipage}{5.5cm}\caption{\label{HONEFL}
Extracted $\fl$ from a comparison of the measured reduced cross
section to an extrapolation of $\ft$, displayed as $\fl$ vs.\ $Q^2$ in
two bins of $y$. Circles and stars denote results from two different
extrapolation methods. Systematic errors for points in the same $y$
bin are correlated. The shaded band indicates $\fl$ as calculated from
the result of a DGLAP fit to $\ft$, the solid line the upper limit
$\fl=\ft$ (the lower limit is $\fl=0$).}\end{minipage}

\end{figure}

The most direct method to measure $\fl$ requires a variation of the
beam energies, which has not been done yet. However, $\fl$ can be
extracted from a comparison of the measured reduced cross section
$\sigma_r$ and an assumption about the behavior of $\ft$, which is
derived, for example, from a DGLAP fit to $\sigma_r$ in the region of
small $y$ (where $\sigma_r\approx\ft$, because the influence of $\fl$
is negligible), and subsequent extrapolation to higher values of $y$. 
Then, $\fl$ is obtained from $\fl= {Y_+}/{y^2}(\ft^\mathrm{fit}
-\sigma_r^\mathrm{measured})$.

Figure \ref{HONEFL} shows the result of such an extraction of $\fl$ by
the H1 collaboration. A measurement of $\fl$ allows to make another
consistency check of the DGLAP formalism, by comparing the measured
$\fl$ to that predicted from the parton densities in the fit. The
method is experimentally challenging, because the region of high
sensitivity to $\fl$ coincides with a region where backgrounds are
high and the quality of the positron detection degrades. In addition,
it relies on assumptions made in the extrapolation.

\section{\boldmath Total Photon-Proton Cross Sections at Low $Q^2$}

In order to study experimentally the region of very low $Q^2$, i.e.\
significantly below $1\GeV^2$, special detectors for the scattered
positron close to the beam line are necessary. The ZEUS detector was
upgraded by the installation of the Beam Pipe Calorimeter (BPC) in
1995 and the Beam Pipe Tracker (BPT) in 1997, a small electromagnetic
sampling calorimeter and a tracking detector of two silicon microstrip
planes, detecting scattered positrons at scattering angles as low as
$1.1^\mathrm{o}$--$1.9^\mathrm{o}$, thus making the region
$0.045\GeV^2<Q^2<0.65\GeV^2$ accessible. H1 installed the VLQ (Very
Low $Q^2$) detector, a device very similar to BPC and BPT, in 1998,
and is working on the analysis of the data.

\begin{figure}[tb] 

\begin{minipage}{5.9cm}

\epsfig{figure=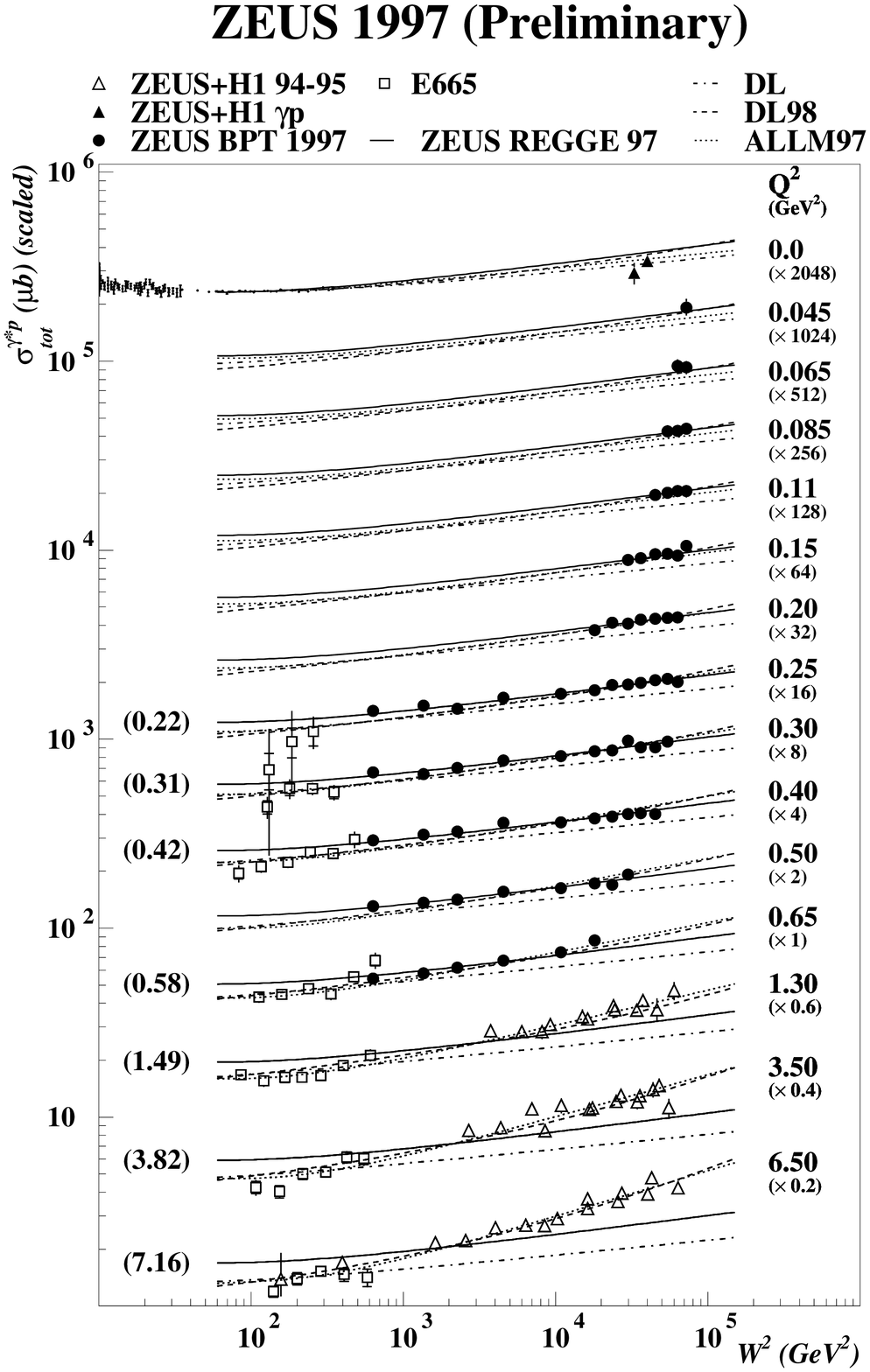,width=5.9cm}


\end{minipage}\hfill\begin{minipage}{5.9cm}

\epsfig{figure=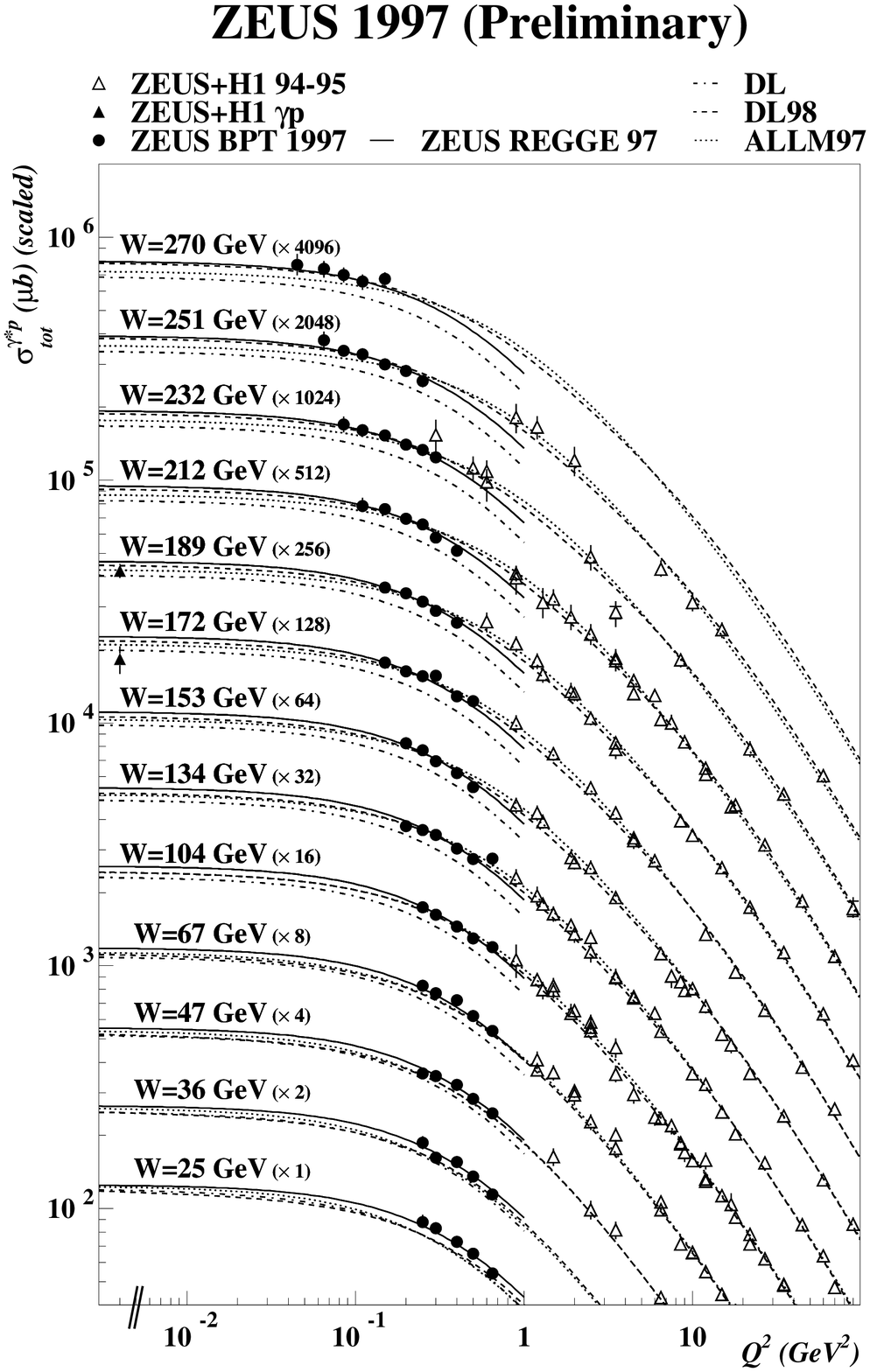,width=5.9cm}

\end{minipage}

\caption{\label{BPC}Measured values of $\sgsp$, plotted vs.\ $W^2$
(left) respectively $Q^2$ (right). Filled circles denote results from
the ZEUS analysis of BPC/BPT data, filled triangles are HERA
measurements at $Q^2=0$, open triangles represent measurements at
higher $Q^2$, and squares denote results from E665. The data are
compared to the parameterizations DL, DL98, and ALLM97, as well as to
a ZEUS Regge fit of the measured data.}

\end{figure}

In Fig.\ \ref{BPC} the ZEUS results of the $\ft$ measurement with the
BPC and BPT, converted to total virtual photon-proton cross sections,
are shown. The data are compared to two parameterizations by Donnachie
and Landshoff (DL \protect\cite{DL94}, DL98 \protect\cite{DL98}), to
ALLM97 \protect\cite{ALLM97} and to a ZEUS Regge fit
\protect\cite{ZEUSPHENO} of the measured data which resembles DL. The
rise of $\sgsp$ with $W^2$ is compatible with a soft behavior for
$Q^2$ up to $0.5\GeV^2$, and becomes steeper at higher $Q^2$. At the
low $Q^2$ values of this analysis, $\sgsp$ is nearly independent of
$Q^2$ (in contrast, it falls like $1/Q^2$ at higher $Q^2$). This can
be exploited to constrain the extrapolation $\sgsp\to\sgp$.

\begin{figure}[tb] 

\begin{minipage}{6cm}

\epsfig{figure=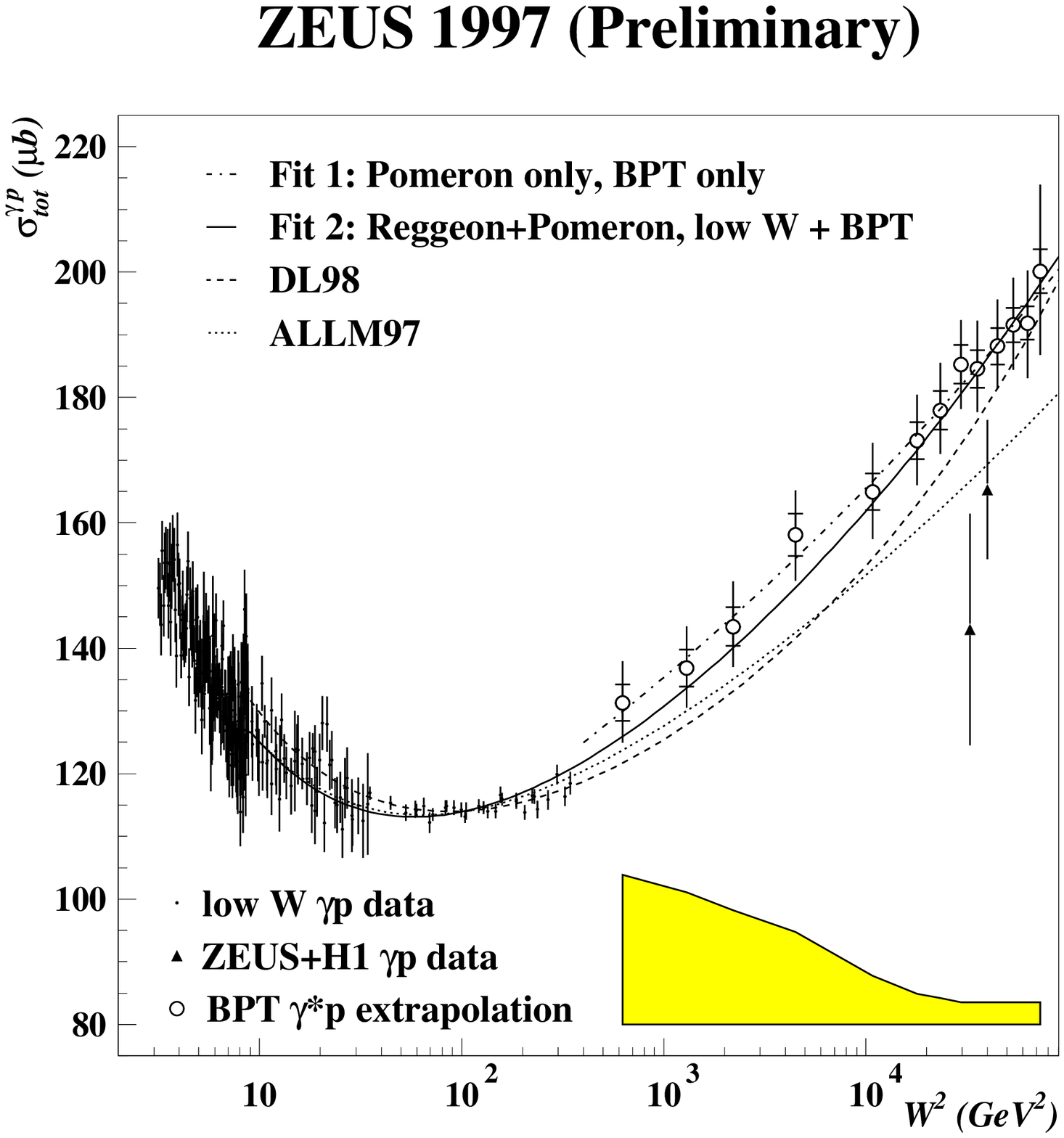,width=6cm} 


\end{minipage}\hfill\begin{minipage}{5.5cm}\caption{\label{BPCEX}
Extrapolated $\sgp$ versus $W^2$ from the ZEUS analysis of BPC/BPT
data, compared to the two direct measurements from ZEUS and H1 and to
data at lower energies. The lines denote two Regge-type fits to the
data, and the DL98 and ALLM97 parameterizations. The shaded band is an
estimate of the magnitude of the model-dependence due to the assumed
$Q^2$ dependence.}\end{minipage}

\end{figure}

In order to extrapolate the measured values of $\sgsp$ to $Q^2=0$, an
assumption about their $Q^2$ dependence is taken from the continuum
part of the GVDM \cite{GVDM} prediction on $\sigma_T$:
$\sgsp(W^2,Q^2)={m_0^2}/{(m_0^2+Q^2)}\sgp(W^2)$ (fitting instead both
the $\sigma_T$ and $\sigma_L$ terms changes the extrapolated values
only within their statistical errors).

The extrapolations are compared to the two direct measurements from
ZEUS and H1 and to data at lower energies in Fig.\ \ref{BPCEX},
together with two Regge-type fits of the form $\sgp(W^2)=A_\reg
W^{2(\alpha_\reg-1)}+A_\pom W^{2(\alpha_\pom-1)}$, and with the DL98
and ALLM97 parameterizations. The extrapolated cross sections have a
tendency to be slightly higher than the directly measured ones. The
model-dependence is estimated from extrapolating with the $Q^2$
dependence of DL, DL98 and ALLM97 (instead of GVDM), and found to be
quite large at small values of $W^2$.

\section{Conclusions}

The HERA collaborations ZEUS and H1 have delivered a wealth of precise
data on the proton structure function $\ft$. At $Q^2>1\GeV^2$, the
data can be described in a self-consistent way by DGLAP fits. At $Q^2$
values below $1\GeV^2$, the transition region from the perturbative to
the non-perturbative regime has been mapped. Regge-inspired models are
used to describe the data here.

\section*{Acknowledgments} I would like to thank the organizers for a
conference full of interesting experiences. It is a pleasure to
acknowledge the outstanding effort of many colleagues in the H1 and
ZEUS collaborations who contributed to the results presented here. I
thank E.~Hilger and B.~Foster for careful reading of the manuscript.
This work was supported by a grant from the Bundesministerium f\"ur
Wissenschaft und Forschung in Germany.


\section*{References} \begin{thebibliography}{99}

\bibitem{DL94} A.~Donnachie and P.V.~Landshoff, \zpc{61} (1994) 139.

\bibitem{DGLAP} V.N.~Gribov and L.N.~Lipatov, \sjnp{15} (1972) 438;\\
L.N.~Lipatov, \sjnp{20} (1975) 96;\\ G.~Altarelli and G.~Parisi,
\npb{126} (1977) 298;\\ Y.L.~Dokshitzer, \spj{46} (1977) 641.

\bibitem{DL98} A.~Donnachie and P.V.~Landshoff, \plb{437} (1998) 408.

\bibitem{ALLM97} H.~Abramowicz and A.~Levy, \dy 97-251.

\bibitem{ZEUSPHENO} ZEUS Collaboration, J.~Breitweg et al., \epjc{7}
(1999) 609.

\bibitem{GVDM} J.J.~Sakurai and D.~Schildknecht, \plb{40} (1972) 121.

\end{thebibliography}
\end{document}